\UseRawInputEncoding 
 \documentclass[11pt, a4paper]{article}

\usepackage{graphicx}
\usepackage{url}
\usepackage{natbib}
\usepackage[utf8]{inputenc}
\usepackage[english]{babel}

\begin{document}

\vspace*{0.35in}

\begin{flushleft}
{\Large
\textbf\newline{Generation possibility of gamma-ray glows induced by photonuclear reactions}
}
\newline
\\
G. Diniz\textsuperscript{1,*},
I.S. Ferreira\textsuperscript{2},
Y. Wada\textsuperscript{1},
T. Enoto\textsuperscript{1},
\\
\bigskip
\bf{1} Extreme Natural Phenomena RIKEN Hakubi Research Team, RIKEN Cluster for Pioneering Research, 2-1 Hirosawa, Wako, Saitama 351-0198, Japan
\\
\bf{2} Instituto de F\'isica, Universidade de Bras\'ilia, Brazil
\\
\bigskip
* gabriel.diniz@riken.jp

\end{flushleft}

%
%


%
%
%
\begin{itemize}
\item Photonuclear reactions triggered by a TGF will sustain a gamma-ray emission level that can promote new RREAs.
\item Both photonuclear reactions and their induced RREA can extend TGF effects to minute timescale.
\item We evaluate the possibility of gamma-ray glows promoted by RREAs originating from photonuclear-induced nuclides decays.
\item An edited version of this paper was published by AGU. Copyright 2021 American Geophysical Union, \cite{diniz2021}.
\end{itemize}

\begin{abstract}
Relativistic runaway electron avalanches (RREAs) imply a large multiplication of high energy electrons ($\sim$1~MeV). Two factors are necessary for this phenomenon: a high electric field sustained over a large distance and an energetic particle to serve as a seed. The former sustains particle energies as they keep colliding and lose energy randomly; and the latter serves as a multiplication starting point that promotes avalanches. RREA is usually connected to both terrestrial gamma-ray flashes (TGFs) and gamma-ray glows (also known as Thunderstorm Ground Enhancement (TGE) when detected at ground level) as possible generation mechanism of both events, but the current knowledge does not provide a clear relationship between these events (TGF and TGE), beyond their possible common source mechanism, still as they have different characteristics. In particular, their timescales differ by several orders of magnitude. This work shows that chain reactions by TGF byproducts can continue for the timescale of gamma-ray glows and even provide energetic particles as seeds for RREAs of gamma-ray glows.
\end{abstract}

\section{Introduction}

High-energy atmospheric phenomena (HEAP) \citep{bbabich2003}, although the late discovery with terrestrial gamma-ray flashes (TGFs: \citet{fishman1994discovery}), have been studied since much earlier as in \citet{wilson1924,wilson1925acceleration,libby1973production}. Different HEAP have been observed over the decades as; spatial distribution and energy spectrum of TGFs \citep{briggs10,Smith2005,tavani2011}), particle production -- in particular neutrons \citep{shah1985neutron, babich2007terrestrial,shyam1999observation,bratolyubova-tsulukidze2004thunderstorms,martin2010observation,rutjes2017neutron, chilingarian2012neutron, chilingarian2012remarks} with the detections reviewed by \cite{babich2019} , as well as extended gamma-ray emissions so-called gamma-ray glows, thunderstorm ground enhancements (TGEs) or long bursts \citep{torii2002observation,tsuchiya2007detection,tsuchiya2012observation,chilingarian2013thunderstorm,kelley2015relativistic,wada2019}. Although \cite{libby1973production} thought the neutron production as a result from nuclear fusion, photonuclear reactions has been proved to be the generation mechanism \citep{babich2006generation, babich2007neutron,babich2007origin}.

\begin{figure}[ht!]
\includegraphics[width=1\textwidth]{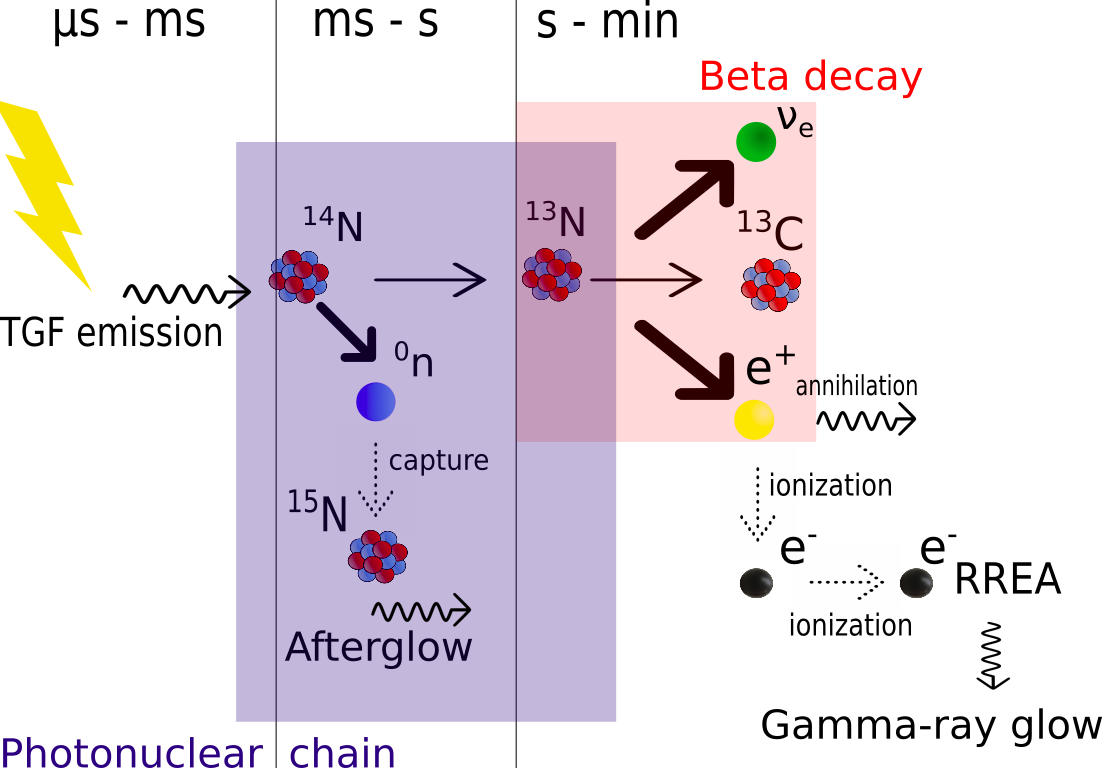}
\caption{\label{fig:ill_time} A distribution of HEAP durations.}
\end{figure}

It has been recently predicted that a TGF can extend its duration \citep{rutjes2017tgf} and confirmed as TGF afterglows which are due to captures of neutrons produced by a TGF in the air \citep{enoto2017neutron}, an earlier gamma generation mechanism by neutrons was proposed by \cite{paiva2013} but through fusion channels. These neutrons, as they are generated, leave unstable atoms as a byproduct which leads to $\beta$-decay \citep{bowers2017gamma,enoto2017neutron, Enoto-2019} and proton emissions \citep{babich2017radio,babich2017nature,babich2019a}. The current whole set of HEAP can then be extended from $\mu s$ timescales with TGFs \citep{fishman1994discovery} up to minutes or hours with the gamma-ray glows \citep{tsuchiya2012observation}, which are summarized in Figure~\ref{fig:ill_time}.

Although the connection from TGFs to neutron emissions and TGF afterglows has recently become clear, there is no record of TGFs followed by gamma-ray glows while \cite{wada2019} reported the latter preceding the former. Both events are thought to be generated by relativistic runaway electron avalanches (RREAs), the feedback mechanism \citep{gurevich1992runaway,dwyer2003fundamental, DWYER12}, and their subsequent \textit{bremstrahlung} \citep{babich04r}. However, they are still seen as separate events.

Nevertheless, the recent $\beta$-decay observations by \cite{umemoto2016,enoto2017neutron} provides high-energy particles that may serve as seeds for RREA, thus generate gamma-ray glows as the atomic-decay emissions continue for a timescale of minutes, and provide a bridge between the phenomena as they are indistinctly correlated with TGF emissions. The possible products and byproducts by nuclear interactions of HEAP are explored further on.

\subsection{The atomic decay possibilities}

There are several photonuclear byproducts \citep{ortega2020,babich2017radio,babich2017nature,babich2019a}; some of them are stable and others are unstable \citep{varlamov1999atlas,dietrich1988atlas}. We focus on the main atmospheric components. Nuclear reactions of our interests are basically the giant dipole resonance (GDR) and neutron capture. Proton-related captures are not considered here since the cross-section of proton captures with nitrogen $^{14}$N$(p,\gamma)^{15}$O peaks at $\sim$ 260~keV with $\sim 10^{-5}$ barns \citep{daigle2013}: the minimum cross-section of neutron captures $^{14}$N$(n,\gamma)^{15}$N is above $10^{-5}$ barns at $\sim$ 20~MeV and, that at 260~keV is $\sim$ $10^{-4}$ barns\footnote{\url{https://www-nds.iaea.org/exfor/endf.htm}}. Also as the proton thermalizes, it can recombine into hydrogen with the available electrons.

Table~\ref{tab:isotope} summarizes all interested photonuclear products and byproducts together with the possibility of $\beta$-decay and Q being the available energy for the decay byproducts.

\begin{table}[h!]
\centering
\caption{Summary of nuclear reaction byproducts with the nuclei of the air species. A gamma as incident particle and Q being the available energy}
\label{tab:isotope}
\begin{tabular}{c|c|c|c|c}
\hline
Target nucleus	& Daughter isotope	& Decay type	& Decay time (min)	& Q (MeV) 		\\ \hline
\multicolumn{5}{c}{Neutron emission}											\\ \hline
$^{14}$N		& $^{13}$N		& $\beta +$	& $\sim 9.965$		& 1.203			\\ \hline
$^{16}$O		& $^{15}$O		& $\beta +$	& $\sim 2.037$		& 1.732			\\ \hline
$^{40}$Ar		& $^{39}$Ar		& $\beta -$	& > years			& 0.054			\\ \hline
\multicolumn{5}{c}{Proton emission}												\\ \hline
$^{14}$N		& $^{13}$C		& Stable		& -				& -				\\ \hline
$^{16}$O		& $^{15}$N		& Stable		& -				& -				\\ \hline
$^{40}$Ar		& $^{39}$C		& $\beta -$	& $\sim 56.2$		& 2.931			\\ \hline
\multicolumn{5}{c}{Neutron capture}												\\ \hline
$^{14}$N		& $^{15}$N		& Stable		& -				& -				\\ \hline
$^{16}$O		& $^{17}$O		& Stable		& -				& -				\\ \hline
$^{40}$Ar		& $^{41}$Ar		& $\beta -$	& $\sim 100.2$		& 1.981			\\ \hline
\multicolumn{5}{c}{Charged-particle production}										\\ \hline
$^{14}$N		& $^{14}$C		& $\beta -$	& \textgreater years	& 0.146			\\ \hline
$^{16}$O		& $^{16}$N		& $\beta -$	& $\sim 0.12$		& 10.420			\\ \hline
$^{40}$Ar		& $^{40}$Cl		& $\beta -$	& $\sim 1.35$		& 7.482			\\ \hline
\end{tabular}
\end{table}

\subsection{This work}
The present paper investigates the minute-lasting effects of a TGF. It is further divided into three sections; Section~\ref{sec:estimate} presents theoretical calculations that shows minute-long TGF effects through its byproducts and chain reactions. Section~\ref{sec:simulations} shows that the $\beta^{+}$ particles have an RREA-generating capability given their isotropic emission. Section~\ref{sec:requirements} discusses the requirements for gamma-ray glows driven by unstable nuclei decay to exist after TGFs.

\section{Theoretical estimates for the TGF-induced glows}\label{sec:estimate}

Currently, two mechanisms are thought of as gamma-ray glow sources: the RREA \citep{gurevich1992runaway} and Modification Of the energy Spectra (MOS) \citep{CHILINGARIAN20121}. RREA depends on electric fields higher than 285 kVm$^{-1}$ \citep{dwyer2003fundamental,babich2004} and an arbitrary seed source (which can be cosmic-rays), while MOS is the cosmic ray energy spectra changed by thunderstorm electric fields and it can occur in lower electric fields and is easier to sustain than RREA. Since MOS is strictly connected to cosmic-rays, TGF-induced glows should happen through the RREA mechanism started by TGF byproducts. Therefore, the byproducts' effects must have minute-long duration and energy enough to trigger an avalanche. Along with these requirements on the source, there is also ambient conditions that must be matched, explored in Section~\ref{sec:requirements}.

This section focuses on phenomena durations and models accordingly to the creation and destruction mechanisms. The timescale analysis shows that TGF products continue for tens of minutes with energies around MeV range, i.e., capable of generating high-energy photon emissions throughout the characteristic time of gamma-ray glows. The following subsections consider rates at standard temperature and pressure (STP) in the atmosphere as is done by \cite{rutjes2017tgf} for simplicity because of all the process frequencies scale in the same fashion with density \citep{Choi07}. The last subsection characterizes the $\beta^{+}$ particles emitted by the unstable nuclei as possible RREA seeds.

The TGF pulse shape is modeled as a gaussian, following \cite{briggs10}, with values to match their observations while we emulate the events after the TGF as non-gaussian pulses in the form $P(t)=n_{\rm p}(1-e^{-t\kappa_{\rm xr}})e^{-t\kappa_{\rm xd}}$. The latter form is characterized by growth and decay rates $\kappa_{\rm x}$, where the sub-index designates the physical process. Hence, the number of particles in a given time depends on creation rate, $\kappa_{\rm xr}$, decreasing rate of, $\kappa_{\rm xd}$, and the parent particle number $n_{\rm p}$. Thus if  $\kappa_{\rm xr} >> \kappa_{\rm xd}$, the particles are created too quickly in comparison to their sink process and would generate a long emission behavior.

\subsection{TGF}\label{sec:TGF}
The TGF spectrum is modeled as Equation~\ref{eq:tgf_spec} \citep{DWYER12}, with $\varepsilon_{\rm th}$ as 7.33~MeV \citep{briggs10} and allows energies up to tens of MeV. One issue is that the spectrum diverges by approaching zero energy. Thus, it is impossible to normalize it without a lower energy cutoff $\varepsilon_{\rm cut}$. There is physically no problem with this divergence because there is no sense of zero energy photons. This energy cutoff needs to be low in comparison with the context and simulations usually implement it as tens of keV \citep{RUTJES16}.

The interesting energies for neutron production are above 10.5~MeV, range in which the photonuclear reactions are relevant \citep{baldwin1947photo}. The number of photons $n_{\gamma}$, for our purposes on probability, in the following calculations is taken as 1 so our calculations are per photon. But, for particle number calculation, \citet{gjesteland2015observation} estimates $10^{17}$-$10^{20}$ photons with energy above 1~MeV,

\begin{equation}
F_{\rm tgf}(\varepsilon) = \frac{n_{\gamma}e^{\frac{-\varepsilon}{\varepsilon_{\rm th}}}}{\varepsilon},
\label{eq:tgf_spec}
\end{equation}

\begin{equation}
    \frac{\int^{\infty}_{10.5}F_{\rm tgf}(\varepsilon)d\varepsilon}{\int^{\infty}_{\varepsilon_{\rm cut}}F_{\rm tgf}(\varepsilon)d\varepsilon} = \Upsilon;
    \label{eq:tgf_GDR}
\end{equation}
considering \cite{gjesteland2015observation} and $\varepsilon_{\rm cut}$ as 1 MeV, $\Upsilon \approx 7\%$ and the total amount of gammas above 10.5~MeV will be 7$\times 10^{15}$-7$\times 10^{18}$. According to \cite{babich2010localization} results of 4.3$\times 10^{-3}$ produced neutrons per photon above 10.5~MeV, there will be approximately  3$\times 10^{13}$ - 3$\times 10^{16}$ neutrons produced in a TGF.

The TGF gaussian (Equation~\ref{eq:tgf_pulse}) parameters are $\mu_{\rm tgf}$ = 130~$\mu s$ and $\sigma_{\rm tgf}$ = 100~$\mu s$ to emulate both the rise time and duration.

\begin{equation}
    P_{\rm tgf}(t) = n_{\gamma}\frac{e^{\frac{-(t - \mu_{\rm tgf})^{2}}{2\sigma_{\rm tgf}^{2}}}}{\sqrt{2\pi \sigma_{\rm tgf}^{2}}}
    \label{eq:tgf_pulse}
\end{equation}

\subsection{Neutron burst}\label{sec:TNB}

As soon as the TGF starts, the photons may interact with the air particles accordingly based on the process probability within their energy range \citep{kohn2017production}. Consequently, the time scale of neutron bursts is basically defined by the frequency of photonuclear reactions $\kappa_{\rm ph-nuc}$ as well as that of neutron captures $\kappa_{\rm capt}$ which is taken to be constant since $\sigma_{\rm capture}\propto (\sqrt{\varepsilon_{\rm neutron})
^{-1}} \propto (v_{\rm neutron})^{-1}$ and $\kappa_{\rm capt}=v_{\rm neutron}\sigma_{\rm capture}n_{\rm air}\propto\frac{n}{n_0}$ as we are using homogeneous density \citep{rutjes2017tgf,blatt1979theoretical}. Both collision frequencies model the neutron pulse shown in Equation~\ref{eq:neu_pulse},

\begin{equation}
P_{\rm neut}(t) \approx n_{\gamma}\Upsilon\frac{\kappa_{\rm ph-nuc}}{\kappa_{\rm ph-absp}}(1- e^{-\kappa_{\rm ph-nuc}t})e^{-\kappa_{\rm capt}t},
    \label{eq:neu_pulse}
\end{equation}

here $n_{\gamma}\Upsilon$ takes into account only TGF photons with energy above 10.5~eV. But these photons may undergo other interactions such as Compton scattering and pair production. Photonuclear reactions are rarer than those \citep{kohn2015calculation}. Such factor agrees with the rate between the photonuclear reaction frequency ($\kappa_{\rm ph-nuc}$) and the photo-absorption frequency ($\kappa_{\rm ph-absp}$) \citep{rutjes2017tgf} with respective values at STP of $\approx 8\times 10^{2}$~s$^{-1}$ and $\approx 2\times 10^{5}$~s$^{-1}$ while the neutron capture frequency is $\approx 15.8$~s$^{-1}$ \citep{rutjes2017tgf,Choi07}.

\subsection{Afterglow}\label{sec:afterglow}

As the neutrons are captured, gamma-rays will be emitted during the capture, the so-called TGF afterglow  \citep{rutjes2017tgf,enoto2017neutron,diniz18}. Thus, the photon number will increase accordingly with both photonuclear and neutron-capture processes. Since $\kappa_{\rm ph-absp} \gg \kappa_{\rm ph-nuc} > \kappa_{\rm capt}$, the photons will be rapidly absorbed by atmospheric particles as they are being created, hence the afterglow time pulse has the same behavior as the neutron's one, but its intensity decreases by a factor $10^{5}$ due to their absorption, resulting in the factor $\frac{\kappa_{\rm ph-nuc}\kappa_{n-\gamma}}{\kappa_{\rm ph-absp}^{2}}$, in which the radiative capture rate is $\kappa_{\rm n-\gamma} = 0.7$~s$^{-1}$ since it is not only possible capture process \citep{rutjes2017tgf,Choi07} resulting in Equation~\ref{eq:aft_pulse},

\begin{equation}
P_{\rm aft}(t) \approx n_{\gamma}\Upsilon\frac{\kappa_{\rm ph-nuc}\kappa_{\rm n-\gamma}}{\kappa_{\rm ph-absp}^{2}}(1- e^{-\kappa_{\rm ph-nuc}t})e^{-\kappa_{\rm capt}t}.
    \label{eq:aft_pulse}
\end{equation}

\subsection{Atomic decay}\label{sec:decay}

Another photonuclear effect is its unstable byproducts confirmed by \citep{dwyer2015,enoto2017neutron, babich2017radio,babich2017nature,kochkin2018,babich2019}. Both nitrogen and oxygen leave, by neutron emission, unstable atoms that will undergo $\beta^{+}$ decay, as indicated in Table~\ref{tab:isotope}. Each unstable atom has the probability to decay with $\kappa_{\rm dc}^{-1}$,  a characteristic time from the moment when it is generated. In consequence of this fact, the decay time pulse (Equation~\ref{eq:dec_pulse}) will start by following the neutron creation and sustain the positron emission during the decay timescale,
\begin{equation}
P_{\rm decay}(t) \approx n_{\gamma}\Upsilon\rho_{\rm rel}\frac{\kappa_{\rm ph-nuc}}{\kappa_{\rm ph-absp}}(1- e^{-\kappa_{\rm ph-nuc}t})e^{-t\kappa_{\rm dc}}.
\label{eq:dec_pulse}
\end{equation}

There are different contributions from nitrogen, oxygen, and even argon. To take this into account in Equation~\ref{eq:dec_pulse}, the relative density $\rho_{\rm rel}$ is introduced. As shown in Table~\ref{tab:isotope}, argon will contribute with $\beta^{-}$ emissions but with a short characteristic rate $\kappa_{\rm dc}$; for this reason and its low relative density together with lower photonuclear cross-section \citep{varlamov1999atlas}, we do not proceed with argon contribution in our estimates.

\begin{figure}[ht!]
\includegraphics[width=1\textwidth]{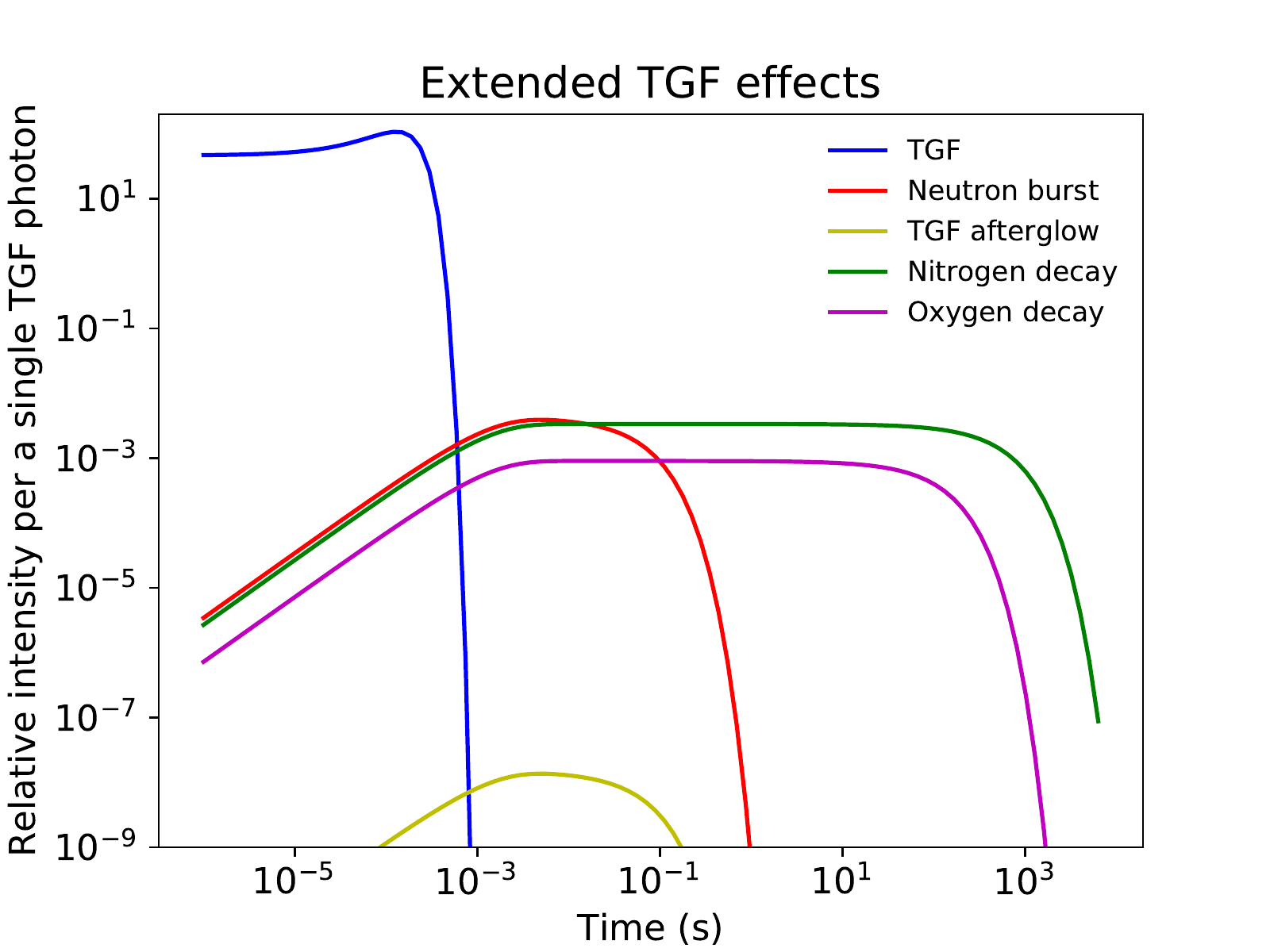}
\caption{\label{fig:pulses}Particle pulses according to equations \ref{eq:tgf_spec}--\ref{eq:dec_pulse}. Following the TGF timeline, up until the nitrogen decay. All curves are thought as per TGF photon with energy between 10 and 30 MeV.}
\end{figure}

%
Although there is no record, found in the literature, of TGFs generating gamma-ray glows, the equation analysis and Figure~\ref{fig:pulses} show that TGF effects reach the time scale of gamma-ray glows as illustrated in Figure~\ref{fig:ill_time}.

\subsection{Energy and temporal scales of $\beta^{+}$-decay emitted positrons}\label{sec:anti}

The temporal estimate makes clear the gamma emission maintenance of a time scale from a TGF to a gamma-ray glow. Depending on the energy, there is the possibility of RREA generation,  possibly together with the feedback mechanism \citep{dwyer2003fundamental,dwyer2012relativistic}. It may be one possible source of gamma-ray glows.

The spectral shape of $\beta$-decay particles can be described by equations \ref{eq:beta_spec}--\ref{eq:Fermi} \citep{wilson68,krane88}, following the assumption of massless neutrinos and high positron energies for simplicity. This simplification does not undermine our estimations since the spectral correction due to the neutrino mass have impact mostly at the extreme spectrum points,
\begin{equation}
    F_{\beta}(\varepsilon) = \sqrt{\varepsilon^{2} + 2\varepsilon m_{e}}(Q - \varepsilon)^{2}(\varepsilon + m_{e})F_{\rm Fermi}(\varepsilon,Z),
    \label{eq:beta_spec}
\end{equation}
\begin{equation}
    S(Z) = \sqrt{1 - \alpha Z^{2}} -1,
    \label{eq:exponent}
\end{equation}
\begin{equation}
    F_{Fermi}(\varepsilon,Z) = \frac{2\pi\nu(\varepsilon,Z)}{1 - e^{-2\pi\nu(\varepsilon,Z})}(\alpha^{2}Z^{2}\omega^{2}(\varepsilon) + 0.25(\omega^{2}(\varepsilon) -1))^{S(Z)},
    \label{eq:Fermi}
\end{equation}
here, $F_{\rm Fermi}(\varepsilon,Z)$ is the Fermi function \citep{krane88}, Z is the daughter (or product) atomic number, $\alpha$ is the fine structure constant, Q is the available energy displayed in Table \ref{tab:isotope}; $\omega$ and $\nu$ are defined respectively as,

\begin{equation}
    \omega(\varepsilon) = \frac{\varepsilon}{m_{e}c^{2}} + 1,
    \label{eq:omega_f}
\end{equation}

\begin{equation}
    \nu(\varepsilon,Z) = \frac{\alpha Z(\varepsilon + m_{e}c^{2})}{\sqrt{\varepsilon^{2} + 2\varepsilon m_{e}c^{2}}},
    \label{eq:nu_f}
\end{equation}

where $m_{e}c^2 = 511$~keV, is the positron rest energy. It is important to note that Equation~\ref{eq:nu_f} diverges as energy decreases, stressing the equation validity only at higher energies.

Equation~\ref{eq:beta_spec} shows the possibilities of kinetic energy for the released positron; and, by energy conservation, during the following electron-positron annihilation the two generated photons will share energy given by $2m_{e}c^{2} + K[e^{+}]$ with $K[e^{+}]$ as the positron kinetic energy and considering the electrons at rest.

\begin{figure}[ht!]
\includegraphics[width=1\textwidth]{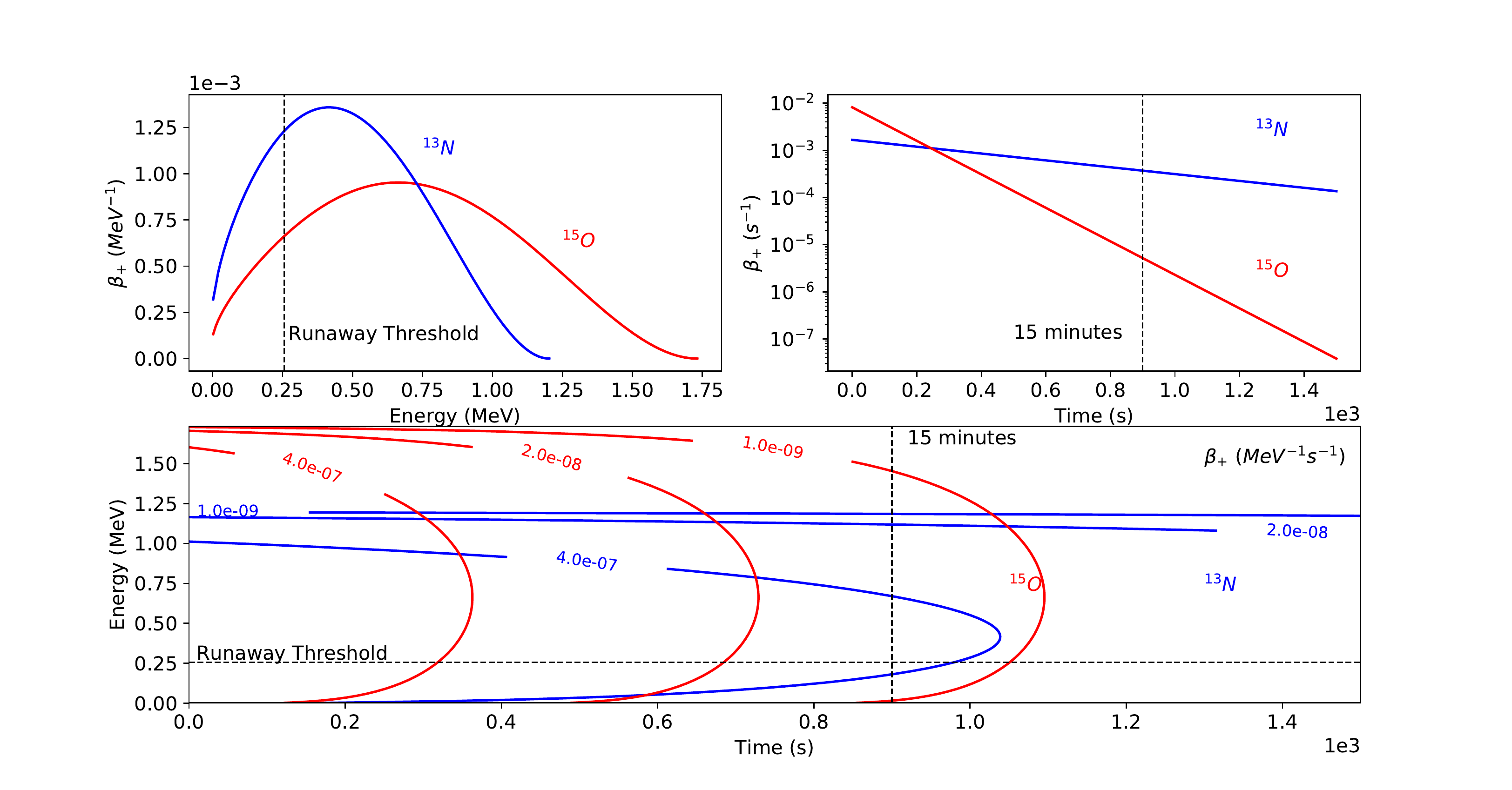}
\caption{\label{fig:prob_beta} $\beta^{+}$ emission as a function of energy (top left panel) and time (top right panel) and the joint information in the bottom panel. The 15-minute line indicates relevant emissions of energetic positron emissions already at the gamma-ray glow time scale. All the $e^{+}$ created with higher energy than the runaway threshold can be an avalanche seed. The runaway threshold line of 0.25 MeV requires an electric field of approximately 0.6 MVm$
^{-1}$ to runaway \citep{kutsyk2012}.}
\end{figure}

As the $\beta$ decay spectrum and its decay in time are statistically independent and normalized, the composed function is the product of both functions. For such, since Equation~\ref{eq:beta_spec} is invalid for lower energies, we normalize it between 2~keV and its maximum value Q. Figure~\ref{fig:prob_beta} displays the spectrum (top left panel), decay pulse (top right panel), and the combined information as a function of time and energy for the $^{13}$N and $^{15}$O decay.

\section{Validating positrons from $\beta^{+}$ decay as RREA seeds by simulations}\label{sec:simulations}
We have performed Monte-Carlo based simulations with GEANT4 software kit focused on the possibility of positron-generating RREAs. GEometry ANd Tracking 4 (GEANT4) \citep{agostinelli2003,allison2006,allison2016} is an open-source toolkit to simulate the particle motion through matter, developed by a collaboration lead by the CERN. It is coded in C++, following an object-oriented methodology. It can simulate the transport of almost all known particles and can include electromagnetic fields. We use the version 10.6 released in December 2019. References and details for these models are presented in the "geant4 Physics reference manual" available at \url{http://geant4.web.cern.ch}. In the present paper, we adapted \cite{SARRIA18} codes to be positron driven but following the same parameters from their supplementary material (available at \url{https://www.geosci-model-dev.net/11/4515/2018/}).

\subsection{Setup of simulations}\label{sec:setup}
The setup is described in detail on the supplementary material of \cite{SARRIA18}. Nevertheless, we implemented simulations with 10$^{4}$ positrons each and electric fields between 0.2 and 1.5~MV~m$^{-1}$. The energies of primary positrons are between 0.02 and 1.73~MeV because the avalanche probability is nearly null for 0.01~MeV \citep{SARRIA18} and the oxygen limiting Q value (see Table~\ref{tab:isotope}). For our purposes, we consider an avalanche as the production of 20 electrons with 1~MeV and use the O4 physics list. Following \cite{SARRIA18}, the avalanche threshold is sufficiently above one runaway electron, a criteria used in earlier works \citep{lehtinen1999monte,li20093d, liu2016adjoint, CHANRION16}, but low enough to deal with computational limitations.

\subsection{Seed influence on RREA possibility}
It is important to highlight the difference between electron avalanche and relativistic runaway electron avalanche. The former can be achieved in a lower energy regime and with electric fields that sustain ionization energy ($\sim$10~eV) producing thermal energy particles. The latter is related to the runaway threshold and, with charge carriers in this energy regime, can be sustained producing relativistic particles \citep{gurevich1992runaway}.

Even though runaway particles require sub-breakdown fields to keep their energy \citep{SKELTVED14}, high electric fields above the breakdown threshold are required to transfer thermal particles into the runaway regime -- the strong-field runaway \citep{Babich1995} or cold runaway \citep{gurevich1961theory}. This change in electric field requirement is explained by the friction curve behavior \citep{peterson1968,MOSS06}, which has a peak at 200~eV for atmospheric air and prevents the thermal particles from transferring to higher energy regimes \citep{diniz19,chanrion2016influence,chanrion2014runaway}.

\begin{figure}[ht!]
\centering
\includegraphics[width=1\textwidth]{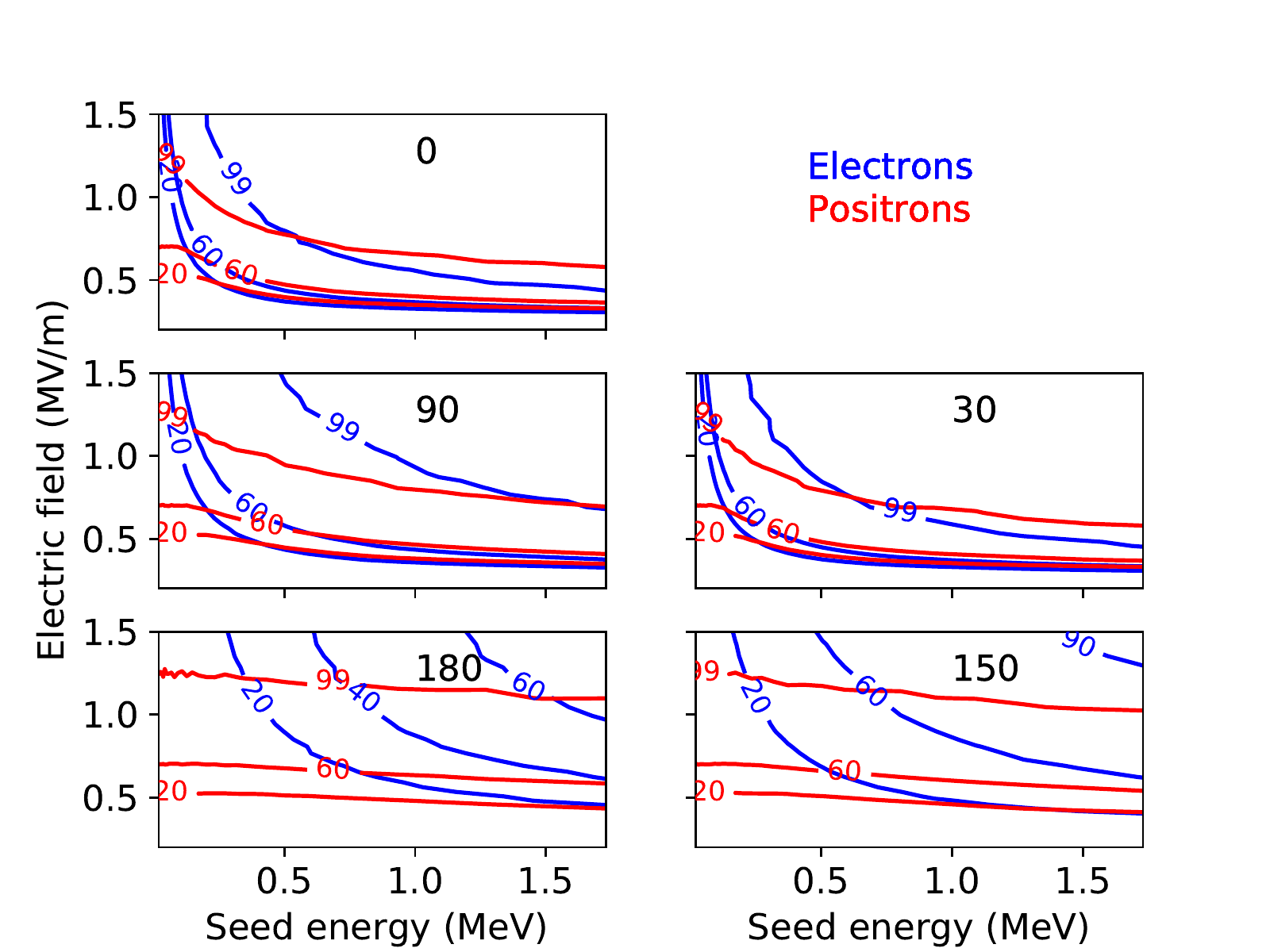}
\caption{\label{fig:prob_rrea}Probability (\%) of an RREA induced by a positron (red) or an electron (blue) with energy given by x-axis and in an electric field in the y-axis. Each panel regards a different angle between the initial velocity and acceleration identified by the black number in degrees at each panel. Here, we consider RREA as the production of 20 electrons with 1~MeV. }
\end{figure}

Our >1~keV energy regime thus requires sub-breakdown electric fields but the field existence is still a necessity for RREAs as will be explored section~\ref{sec:electric}. The electric field serves as an energy source and provides it enough to not only sustain the avalanche but also accelerate the produced particles to keep them in the runaway regime. Figure~\ref{fig:prob_rrea} displays the probability to reach RREA status, here defined as the production of 20 electrons with one~MeV electron/positron, as a function of electric field and primary particle energy.


Figure~\ref{fig:prob_rrea} demonstrates the differences between a positron RREA seed and an electron one. At low energies, the positron seed has a significantly higher chance of avalanche production because, after the seed thermalizes, it will annihilate producing gamma-rays able to produce high-energy electrons while low energy electrons always need to be accelerated to emit gamma-ray.

\subsection{From decay products to RREA}

It is important to notice that while high-energy positrons do not lose all energy to annihilate, they also produce high-energy electrons. Both features, the electron production, and the annihilation make positron RREA seeds more efficient than electron ones.


The decay products will not be necessarily emitted aligned with the electric field. This alignment modifies the RREA production chance, as shown by Figure~\ref{fig:prob_rrea}. Nevertheless, the results display at least 20\% probability of RREA generated by positrons with any considered energy for all simulated initial alignment with electric fields $\geq$ 0.5~MVm$^{-1}$.

As RREA can promote multiple HEAP events, this possibility permits a cyclic view on the phenomena as they can take origin through the byproducts of other events. There are limits for such a cycle \citep{dwyer2007relativistic}: the possible chain reaction of the high-energy particle production is limited by the need for an electric field, thus it is limited in time by the thunderstorm duration and in space by the electric field decreasing with distance.

\section{Gamma-ray glow requirements}\label{sec:requirements}

Two conditions are necessary for gamma-ray glows: (1) enough seed number and (2) strong and long electric fields. The former concerns the source capability to provide a large number of energetic particles while the latter restricts the ambient electric field to values and sizes capable of sustaining the particle multiplication during the cascade motion through the air.

\subsection{$\beta^{+}$ decay versus cosmic-ray seed}\label{sec:seed}

Cosmic rays are thought to be RREA seed providers \citep{wilson1925acceleration,williams2010origin} and related with gamma-ray glow source \citep{wada2019}. The flux of energetic cosmic-ray electrons represents a reference quantity to compare phenomenological candidates to RREA and gamma-ray glow sources.

\citet{wada2019} measurements occurred at (36.5$^{\circ}$N, 136.6$^{\circ}$E) on 9 January 2018, during a winter thunderstorm in Japan. The energetic cosmic-ray shower spectra (differential flux) can be retrieved for such coordinates using EXPACS \citep{sato2015,sato2016} model\footnote{\url{http://phits.jaea.go.jp/expacs/}} in units of [/MeV/cm$^2$/s]. We have summed the differential flux of electrons and positrons to compare with the $\beta^{+}$-decay seed deposit.

The radial density distribution of secondary cosmic-ray electrons can be described by the well-known NKG empirical formulae which depend on a characteristic radius scale (R $\approx$ 115~m for air) \citep{gaisser2016, GUREVICH1999}. We estimate the cosmic ray shower spectra integrated over the area by multiplying EXPACS results by the orthogonal circular area with a 115~m radius.

This comparison is chosen because while the secondary cosmic ray particles all reach the thunderstorm region from upwards and are distributed radially, the unstable nuclei are left from photonuclear reactions in a longitudinal developed fashion and the $\beta$ particles are emitted isotropically, not allowing a direct flux comparison.

To compare the cosmic ray secondary spectra in [MeV$^{-1}$s$^{-1}$] with the $\beta$-emission from unstable nuclei, we considered the normalized $\beta$-decay spectra multiplied by the rate of decayed particles. For such, we are considering that all neutrons are produced by the same channel, i.e., single neutron outgoing from the nuclei and leaving unstable $^{13}$N or $^{15}$O.

\begin{figure}[ht!]
\includegraphics[width=1\textwidth]{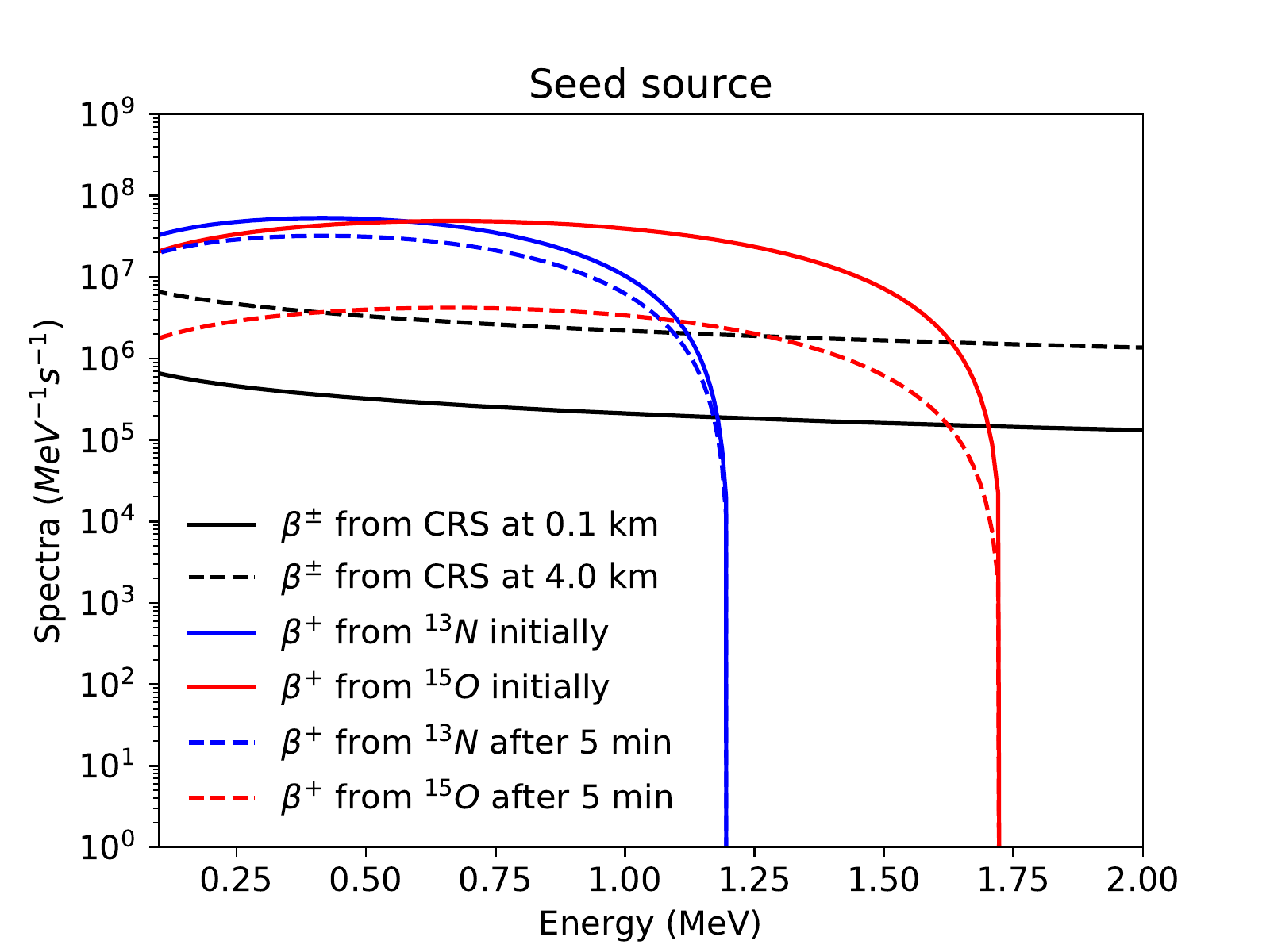}
\caption{\label{fig:CRSbeta} Cosmic-ray secondary spectra compared (electrons and positrons combined) with the $\beta^{+}$-decay spectrum at different moments. The blue (red) lines regards emissions from $^{13}$N ($^{15}$O) right after the neutrons generation (solid line) and after five minutes (dashed lines). The black lines are the summed electron and positron spectra from cosmic ray secondaries at the time and geographic location of \cite{wada2019} detections for two different altitudes: 100 m (solid line) and 4 km (dashed line).}
\end{figure}

Using the total neutron number, 3$\times$10$^{13}$, estimated in section~\ref{sec:TGF} and following our approximations, there will be approximately 2.3$\times 10^{13}$ $^{13}$N and 6.2$\times 10^{12}$ $^{15}$O to decay. Figure~\ref{fig:CRSbeta} compares cosmic ray shower of electrons and positrons with the $\beta$-decay spectra.

The higher number of avalanche seed particles from unstable nuclei than cosmic ray shower background yield for the comparable energy range, hence, making the $\beta^{+}$ emitters a viable enhancer of energetic particle flux for minute-scale after the TGF.

Although the cosmic-ray secondaries cover a wider energy range than the $\beta$-decay, the total rate of particles from the latter is greater than from the former. Integrating the Figure~\ref{fig:CRSbeta} spectra over the energy, the rate of electrons and positrons from cosmic rays is approximately 1.8$\times$10$^{7}$ and 1.8$\times$10$^{6}$ per second at 4~km and 0.1~km altitude respectively while the rate of decayed unstable nuclei is, initially, 3.9$\times$10$^{10}$ decayed $^{13}$N  and 5.1$\times$10$^{10}$ decayed $^{15}$O per second; after 5 minutes the rates will be  2.1$\times$10$^{10}$ decayed $^{13}$N  and 4.4$\times$10$^{9}$ decayed $^{15}$O per second.

\subsection{Thundercloud electric field}\label{sec:electric}

During the motion through the air, energetic particles will lose their energy due to stochastic collisions. Thus, there is a need for a constant energy gain to sustain the avalanche capability of generating gamma-rays. The dynamic friction implements a minimal value for runaway-causing electric fields of 216~kVm$^{-1}$ \citep{gurevich1992runaway, gurevich2001} while the threshold for RREA is approximately 285
~kVm$^{-1}$ \citep{dwyer2003fundamental, babich2004} due to stochastic Coulomb scattering that prevents an optimal energy gain from the electric field.

TGFs are correlated with lightning discharges which will remove charged regions, from the thundercloud, changing the ambient electric field and promoting an abrupt end to high energy radiation \citep{parks81,mccarthy1985further,tsuchiya2007detection,kelley2015relativistic}. Such effect imposes a challenge for a gamma-ray glow following the TGF although the latter produce possible avalanche seed particles. Nevertheless, observations also show long bursts despite lightning occurrence \citep{chilingarian2017} suggesting that not always the discharge will interrupt a long gamma emission; moreover, \cite{CHILINGARIAN2020} reports intermittent TGE that ends with lightning discharge and continuously recover the count rate in minute-scale.

The unstable leftovers from neutron-generating photonuclear reactions have a long duration following the characteristic decay time and will be distributed over the elongated photon path. These two features allow them to feed the ambient with avalanche seeds while the electric field recovers from the lightning discharge and reach different cloud regions that may still have a charge configuration.

\section{Conclusion}
We have shown an extended TGF timescale up to tens of minutes via its particles byproducts. Moreover, the possibility of the TGF particle chain production results in new RREA that can be sustained to gamma-ray glows duration. Although the decay emission is isotropic, our simulations display positrons are capable of inducing RREA efficiently despite their initial orientation regarding the electric field. Due to the $\beta$ decay nature, long-lasting extra supply of energetic particles is available thus, if the thundercloud electric field recovers in minute-scale after the TGF. There will be a chance for a follow up gamma-ray glow despite the associated lightning charge removal.
Such conditions allow HEAP a cyclic nature as they keep producing RREA and there are multiple possible outcomes from it. The need for an electric field introduces the actual limitation to this chain-reaction process as it is limited both in time and space, i.e., the electric field variations can stop the cycle.
This new perspective on HEAP connection explains how a TGF may provide a possibility for a gamma-ray glow to take place and even generate the gamma-ray glow itself depending mostly on the ambient electric field dynamics since, as it is shown, parallel to the constant cosmic ray. The TGF itself provides a long-lasting supply of avalanche seed particles.

\section*{Acknowledgement}
This work and G.D. is supported by MEXT/JSPS KAKENHI Grant No. 19H00683; T.E. and Y.W. are supported by Hakubi Research Project and Special Postdoctoral Researcher fellowship program of RIKEN, respectively; I.F. is supported by CAPES. The simulation results are available online (doi: 10.17632/kv3bnscc8b.4). The codes for the simulations presented here can be found at the supplementary material of Sarria et al. [2018]: https://gmd.copernicus.org/articles/11/4515/2018/gmd-11-4515-2018.html. We sadly inform the community that our coauthor Dr. Ivan Soares Ferreira passed away in October. He was an incredible friend, scientist, and professor. Although not everyone knew him, his decease is a tremendous loss for all the academic community. The two last songs he sent singing were: "São Murungar" by Bezerra da Silva, in his profound love for samba;  and, ironically, "My way" popularized in the voice of Frank Sinatra. May Ivan be always remembered as the great person he was and may we be capable of honor his memory.
\bibliographystyle{unsrtnat}
\bibliography{main}


\end{document}